\documentclass[11pt,a4paper]{article}
\usepackage{jheppub}
\usepackage{longtable}
\usepackage{bm}

\newcommand*{\no}{\noindent}
\newcommand*{\bea}{\begin{eqnarray}}
\newcommand*{\eea}{\end{eqnarray}}
\newcommand*{\be}{\begin{equation}}
\newcommand*{\ee}{\end{equation}}

\newcommand*{\pref}[1]{(\ref{#1})}

\newcommand*{\mn}{{\mu\nu}}

\newcommand*{\nn}{\nonumber}

\newcommand*{\tr}{\mathrm{tr}}

\title{Topological aspects of G$_2$ Yang-Mills theory}

\author[1]{Ernst-Michael Ilgenfritz}\emailAdd{ilgenfri@lhep.jinr.ru}
\author[2]{Axel Maas}\emailAdd{axelmaas@web.de}

\affiliation[1]{Joint Institute for Nuclear Research, VBLHEP, 141980 Dubna, Russia}
\affiliation[2]{Institute for Theoretical Physics, Friedrich-Schiller-University Jena, Max-Wien-Platz 1, D-07743 Jena, Germany}

\abstract{Yang-Mills theory and QCD are well-defined for any Lie group as gauge group. 
The choice G$_2$ is of great interest, as it is the smallest group with trivial center 
and being at the same time accessible to simulations. This theory has been found to have 
many properties in common with SU(3) Yang-Mills theory and QCD, permitting to study the 
role of the center. Herein, these investigations are extended to topological properties 
of G$_2$ Yang-Mills theory. After giving the instanton construction for G$_2$, topological 
lumps with instanton topological charge are identified in cooled lattice configurations. 
The corresponding topological susceptibility is determined in the vacuum and at low and 
high temperatures, showing a significant response to the phase structure of the theory.}


\begin{document}

\maketitle

\section{Introduction}

Yang-Mills theory and QCD are well-defined theories for an arbitrary (semi-)simple Lie 
group as gauge group. One remarkable choice for the group is the exceptional Lie group 
G$_2$ instead of the physical group SU(3).
Since its center is trivial the Wilson confinement criterion is not fulfilled, even in 
the pure Yang-Mills case \cite{Holland:2003jy}. The reason is that any static fundamental 
charge can be screened by three adjoint charges, i.\ e.\ gluons \cite{Holland:2003jy}. 
Nonetheless, the theory has only gauge-invariant bound states as observable 
states \cite{Holland:2003jy}. In fact, the gluons show a behavior quite similar to SU($N$) 
gauge theories \cite{Maas:2005ym,Maas:2007af,Maas:2010qw}.

It is not only in this respect that the G$_2$ case resembles SU($N$) Yang-Mills theory. 
Just like for SU(3) gauge theories it shows a first-order phase transition at finite 
temperature \cite{Pepe:2006er,Greensite:2006sm,Cossu:2007dk}, which 
is accompanied by a quenched chiral transition \cite{Danzer:2008bk}. There are other 
features that make the theory more like a theory with dynamical matter content. Especially 
the screening of static fundamental charges leads to string-breaking, though 
at intermediate distance a linearly rising static quark potential is 
present \cite{Greensite:2006sm,Liptak:2008gx,Wellegehausen:2010ai}, and the theory can be 
described using an effective Polyakov loop dynamics \cite{Wellegehausen:2009rq}. The latter 
fact may be related to an SU(3) subgroup structure \cite{Holland:2003jy}, to which the 
theory can be broken down using the Higgs mechanism \cite{Holland:2003jy,Wellegehausen:2011sc}.

Besides these very interesting conceptual properties of the theory, it 
offers advantages that might also be of practical importance. Since all representations 
of G$_2$ are real, it is possible to simulate it using standard 
importance sampling techniques in the presence of dynamical, fundamental 
quarks \cite{Maas:2012wr} also at finite baryonic density without the notorious sign
problem. It is therefore an interesting test case for model calculations. Since its spectrum 
contains fermionic baryons (besides bosonic baryons) \cite{Holland:2003jy}, it also offers 
qualitative insight into a theory with fermionic bound state degrees of freedom at finite 
density.

Given the similarities between the G$_2$ gauge theory and ordinary QCD and the practical 
usefulness of the G$_2$ case, it is an interesting question, whether topological aspects 
play a similar role as in usual QCD. It has already been argued that the role of vortices 
is modified \cite{Greensite:2006sm} compared to their role in QCD \cite{Greensite:2003bk}. 
The question of monopoles \cite{DiGiacomo:2008nt} and dyons \cite{Diakonov:2010qg} has 
been addressed in principle, showing a similar structure as in ordinary QCD. 
Given the coincidence of the chiral and the Polyakov loop transition in the pure Yang-Mills
case \cite{Danzer:2008bk}, we are interested in this paper whether topological charge 
carriers exist, which could play a similar role in this connection as in ordinary 
QCD \cite{Schafer:1996wv}.

To begin, we will construct the explicit one-instanton solution in the continuum in 
section \ref{scontinuum}. In section \ref{slattice} we will describe our lattice simulations
and cooling procedure used to identify topological lumps. We will discuss the resulting 
structures in section \ref{scharge}. Finally, we will determine the topological charge 
susceptibility in the vacuum and at finite temperature in section \ref{sfinitet}. 
A few concluding remarks will be given in section \ref{ssum}.

\section{G$_2$ instanton solution in the continuum}
\label{scontinuum}

Like the SU($N$) gauge group, G$_2$ supports instanton solutions. 
This can be most easily seen using the McFarlane decomposition \cite{Macfarlane:2002hr} 
of a G$_2$ element $g$ 
\be
g=Z\begin{pmatrix} U & 0 & 0 \cr 0 & 1 & 0 \cr 0 & 0 & U^* \end{pmatrix}=e^{ig^a\tau^a},\nn
\ee
\no where $Z$ is an element of $S^6$ and $U$ is an element of SU(3). Thus, the generators 
$\tau^a$ can also be chosen such that six of them generate elements from the coset and 
the other eight generate the elements from the su(3) algebra, taking the form
\be
\tau_{1...8}=\begin{pmatrix} u_{1...8} & 0 & 0 \cr 0 & 0 & 0 \cr 0 & 0 & -u_{1...8} \end{pmatrix},\nn
\ee
\no where the $u_i$ are the generators of the algebra su(3). 
Given the SU(2) instanton solution $A^\text{SU(2)}_\mu$ \cite{Bohm:2001yx}
\bea
A^\text{SU(2)}_\mu&=&\frac{2}{r^2+\lambda^2}t_\mn r_\nu\nn\\
t_\mn&=&\frac{1}{4i}(t_\mu\bar t_\nu-t_\nu\bar t_\mu)\nn\\
t_\mu&=&(i\vec t,1)\nn\\
\bar t_\mu&=&(-i\vec ,1),\nn
\eea
\no with the Pauli matrices $t_i$, the corresponding G$_2$ gauge field can be obtained, 
with the direct su(3) embedding of the su(2) solution, as
\bea
A_\mu^{\text{G}_2}&=&\begin{pmatrix}A^\text{SU(3)}_\mu & 0 & 0 \cr 0 & 0 & 0 \cr 0 & 0 & -A^{\text{SU(3)}*}_\mu\end{pmatrix}  \nn\\
A_\mu^{\text{SU(3)}}&=&\begin{pmatrix}A^\text{SU(2)}_\mu & 0 \cr 0 & 0\end{pmatrix}\quad\text{and re-distributions.}  \nn
\eea
\no This gauge field solves the the self-duality equations
\be
F_\mn=\frac{1}{2}\epsilon_{\mu\nu\rho\sigma}F_{\rho\sigma}\nn
\ee
\no immediately, due to the subgroup structure of G$_2$, 
whereas the S$^6$ part acts as a spectator. The anti-instanton solution can be constructed 
along the same lines. The most remarkable difference 
compared to the SU(3) or SU(2) case is the topological charge of the G$_2$ instanton
\be
Q=\frac{1}{64\pi^2}\int d^4x\epsilon_{\mu\nu\rho\sigma} F^a_{\mu\nu} F^a_{\rho\sigma}=2\nn,
\ee
\no which is twice as large as the one of the (embedded) SU(2) instanton. Consequently, 
also the corresponding action is twice as large. This result is already sufficient to 
motivate the following numerical studies.
The existence of exact solutions with single instanton topological charge will not be 
addressed here, though the numerical results below 
are suggestive in favor of their existence.

\section{Lattice setup and cooling}
\label{slattice}

\begin{table}
 \begin{tabular}{|c|c|c|c|c|c|c|c|c|c|}
  \hline
  $\beta$ & $N_t$ & $N_s$ & conf. & therm. & sweeps & $a$ [fm] & $L=aN_s$ [fm] & $T=\frac{1}{aN_t}$ [MeV] & $\frac{T}{T_c}$ \cr
  \hline
  9.515 & 8 & 8 & 214 & 380 & 80 & 0.210 & 1.68 & - & - \cr
  9.515 & 12 & 12 & 215 & 420 & 120 & 0.210 & 2.52 & - & - \cr
  9.515 & 16 & 16 & 129 & 460 & 160 & 0.210 & 3.36 & - & - \cr
  \hline
  9.6 & 8 & 8 & 211 & 80 & 380 & 0.170 & 1.36 & - & - \cr
  9.6 & 12 & 12 & 210 & 420 & 120 & 0.170 & 2.04 & - & - \cr
  9.6 & 16 & 16 & 124 & 460 & 160 & 0.170 & 2.72 & - & - \cr
  \hline
  9.73 & 8 & 8 & 226 & 380 & 80 & 0.134 & 1.07 & - & - \cr
  9.73 & 12 & 12 & 149 & 420 & 120 & 0.134 & 1.61 & - & - \cr
  9.73 & 16 & 16 & 132 & 460 & 160 & 0.134 & 2.14 & - & - \cr
  \hline
  \hline
  9.6 & 6 & 12 & 217 & 420 & 120 & 0.170 & 2.04 & 193 & 0.757 \cr
  9.73 & 6 & 12 & 220 & 420 & 120 & 0.134 & 1.61 & 245 & 0.961 \cr
  9.765 & 6 & 12 & 155 & 420 & 120 & 0.129 & 1.55 & 255 & 1.00 \cr
  9.85 & 6 & 12 & 118 & 420 & 120 & 0.119 & 1.43 & 276 & 1.08 \cr
  10 & 6 & 12 & 117 & 420 & 120 & 0.0954 & 1.14 & 344 & 1.35 \cr
  \hline
  \hline
  9.6 & 6 & 16 & 138 & 460 & 160 & 0.170 & 2.72 & 193 & 0.757 \cr
  9.73 & 6 & 16 & 150 & 460 & 160 & 0.134 & 2.14 & 245 & 0.961 \cr
  9.765 & 6 & 16 & 141 & 460 & 160 & 0.129 & 2.06 & 255 & 1.00 \cr
  9.85 & 6 & 16 & 121 & 460 & 160 & 0.119 & 1.90 & 276 & 1.08 \cr
  10 & 6 & 16 & 212 & 460 & 160 & 0.0954 & 1.52 & 344 & 1.35 \cr
  \hline
 \end{tabular}
 \caption{\label{config}List of configurations employed. $N_t$ and $N_s$ are the temporal 
and spatial extension of the lattice. ``Therm.'', ``sweeps'', and ``conf.'' denote the 
number of 
thermalization and decorrelation sweeps, and the number of configurations, respectively. 
The scale has been set using the string-tension values given in \cite{Liptak:2008gx}, 
using the same strategy as in \cite{Danzer:2008bk} and setting the intermediate distance 
string tension equal to $(440$ MeV$)^2$. Note that for the Wilson action the critical 
value of $\beta$ for a time extension of $N_t=6$ is 9.765 ($T_c=255$ MeV), significantly 
above the bulk transition, which occurs at $\beta=9.45$ \cite{Pepe:2006er,Cossu:2007dk}. 
In all cases, many independent runs haven been performed to reduce residual correlations.}
\end{table}

The G$_2$ gauge field configurations for pure Yang-Mills theory used here have been 
generated using a combined overrelaxation and heat-bath algorithm with respect to the
one-plaquette Wilson action
\be
A=\beta\sum_{x;\mu>\nu}\left(1-\frac{1}{7}\tr U_\mn\right)\label{wa}
\ee
\no with the plaquette $U_\mn$, as described in \cite{Maas:2007af}. The list of lattice setups employed to study both zero and finite 
temperature are given in table \ref{config}. To make sure that auto-correlations, very 
prominent in SU($N$) gluodynamics 
\cite{Schaefer:2010hu}, are duly taken into account only few measurements have been 
performed in many individual runs, with a large number of dropped configurations in 
between.

To identify topological structures, we needed to reduce the ultraviolet fluctuations. 
Conventional APE smearing \cite{DeGrand:1997ss} turned out to be rather inefficient 
for G$_2$, requiring too many sweeps to reduce the fluctuations substantially. 
We therefore employed the following cooling algorithm:

In a checker-board fashion, every lattice site is visited. 
The links in all directions at that lattice site are
then modified such as to locally minimize the action, 
which is done by a heat-bath update 
where only proposals reducing or keeping the action are accepted. 
This heat-bath update is performed for a single SU(2) sub-group of a G$_2$ element. 
In order to update all subgroups, after each such sweep a random gauge transformation 
of all links is performed, which mixes the different sub-groups. This cycle is done 
fifteen times before  
a single cooling sweep of the lattice is considered as completed.

On these smeared configurations the measurements have then been performed. 
The observables have been 
the local action density and the local topological charge density $q$. The latter 
has been measured using the simplest lattice realization of the continuum operator
\be
q(x)=\frac{1}{256\pi^2}\tr\epsilon_{\mu\nu\rho\sigma}F_\mn(x) F_{\rho\sigma}(x),\label{q}
\ee
\no by calculating first the field strength tensor $F_\mn$ at site $x$ from the link 
variables, and then calculating the product \pref{q}. The full topological charge $Q$ 
is then obtained by summation
\be
Q=\sum_x q(x)\nn.
\ee
\no We have furthermore determined the topological charge susceptibility
\be
\chi_Q=\frac{1}{N_t N_s^3}(\langle Q^2\rangle-\langle Q\rangle^2)\nn,
\ee
\no which has units of (energy)$^4$.

\section{Cooling histories, topological lumps, and topological charge}
\label{scharge}

\begin{figure}
 \includegraphics[width=0.5\linewidth]{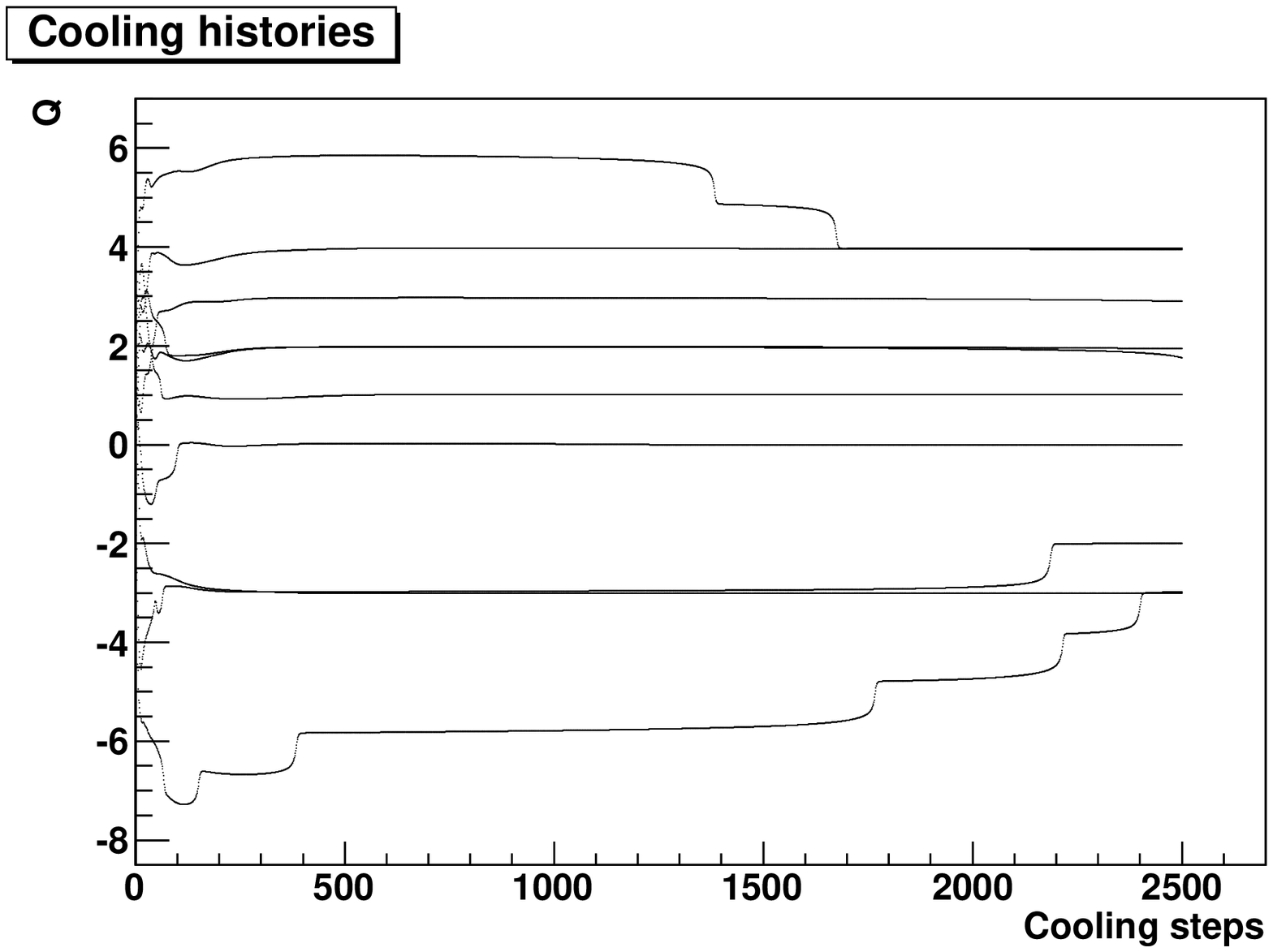}\includegraphics[width=0.5\linewidth]{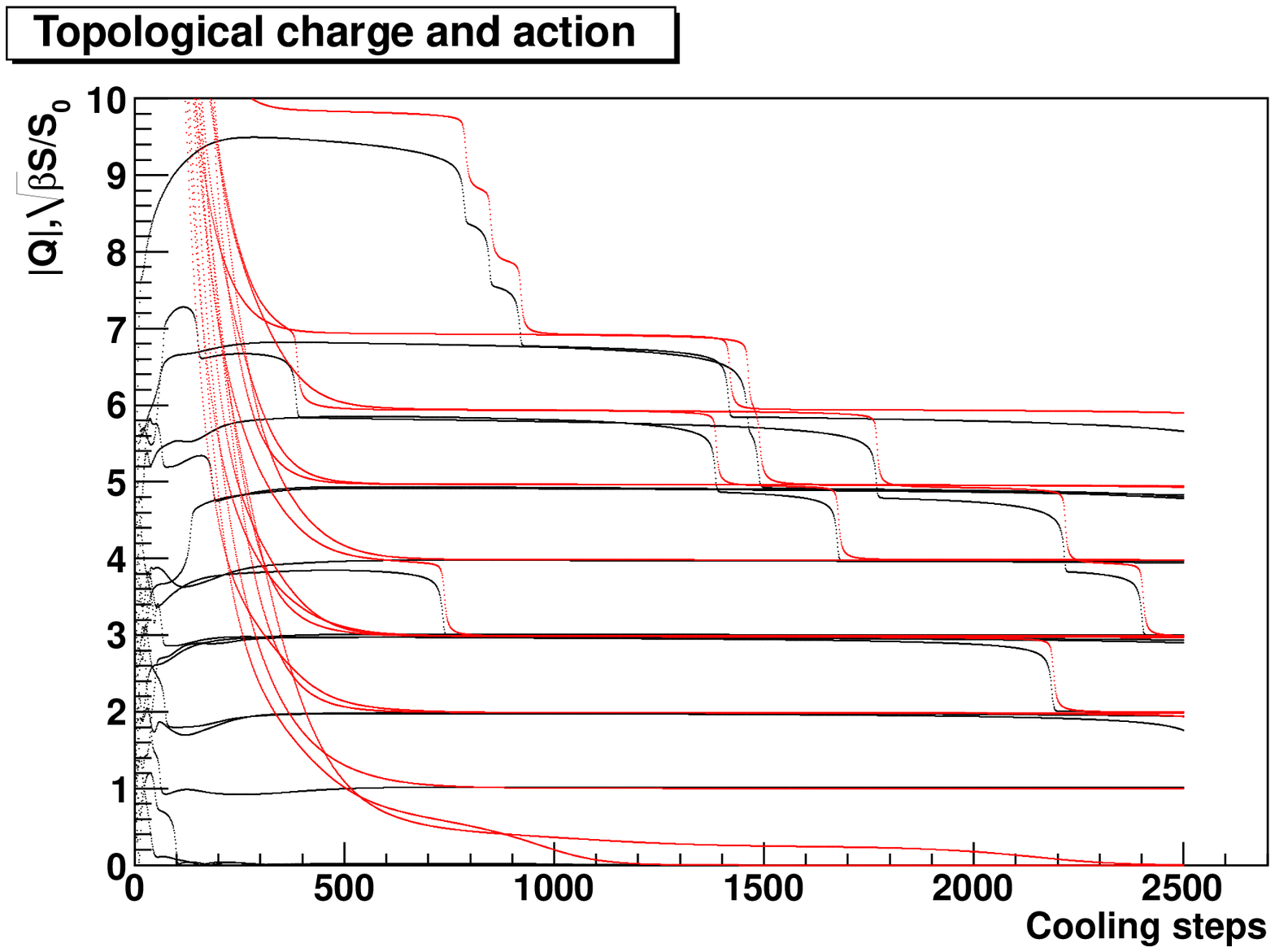}
 \caption{\label{history}Left panel: Cooling histories of the topological charge $Q$ 
for a $12^4$ lattice at $\beta=9.515$. Right panel: the absolute value of $Q$ for the 
same cooling histories (black lines) compared to the cooling curves of the action 
divided by the naive action $S_0=14\pi^2\beta$ of a $Q=1$ object (red lines).}
\end{figure}

Cooling histories for typical configurations are shown in figure \ref{history}. The first 
observation is that the cooling process is still rather inefficient, and requires many 
sweeps to significantly change a configuration. 
The next observation is the appearance of essentially integer-valued plateaus, which are 
very stable in the course of the cooling process, while the changes between the plateaus 
are rather rapidly going on.
This is the typical structure expected for the presence of topologically stable lumps. 
Furthermore, also the full Wilson action, including the constant term $6\beta V$ in \pref{wa}, exhibits the same 
plateau structure in a one-to-one correspondence. Thus, again as in SU($N$) theories, 
the topological charge dominates the action  after a sufficient number of cooling sweeps 
(above roughly 1200). In other words, globally (anti)selfdual configurations are obtained
by cooling.

\begin{figure}
 \includegraphics[width=\linewidth]{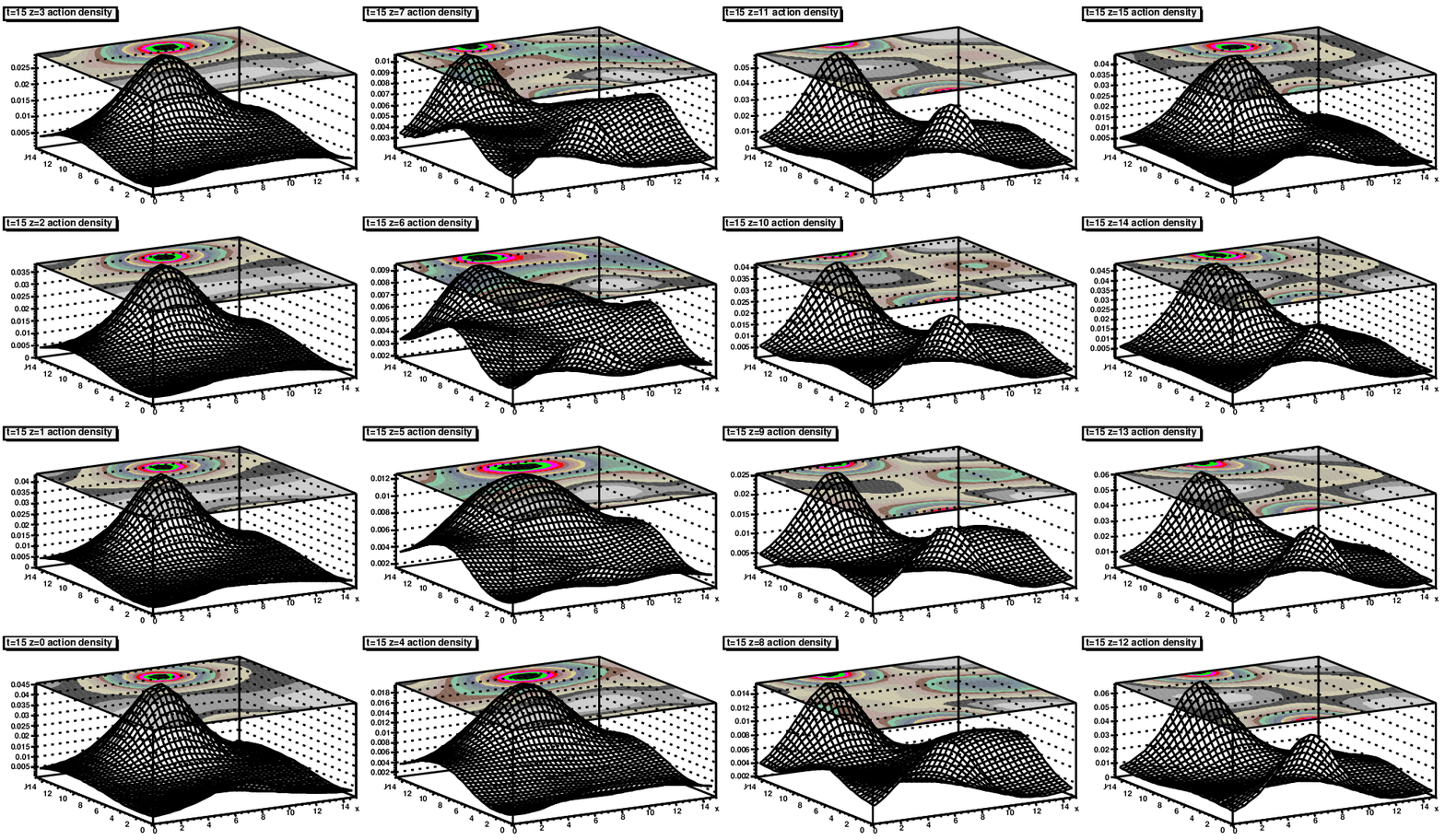}\\
 \includegraphics[width=\linewidth]{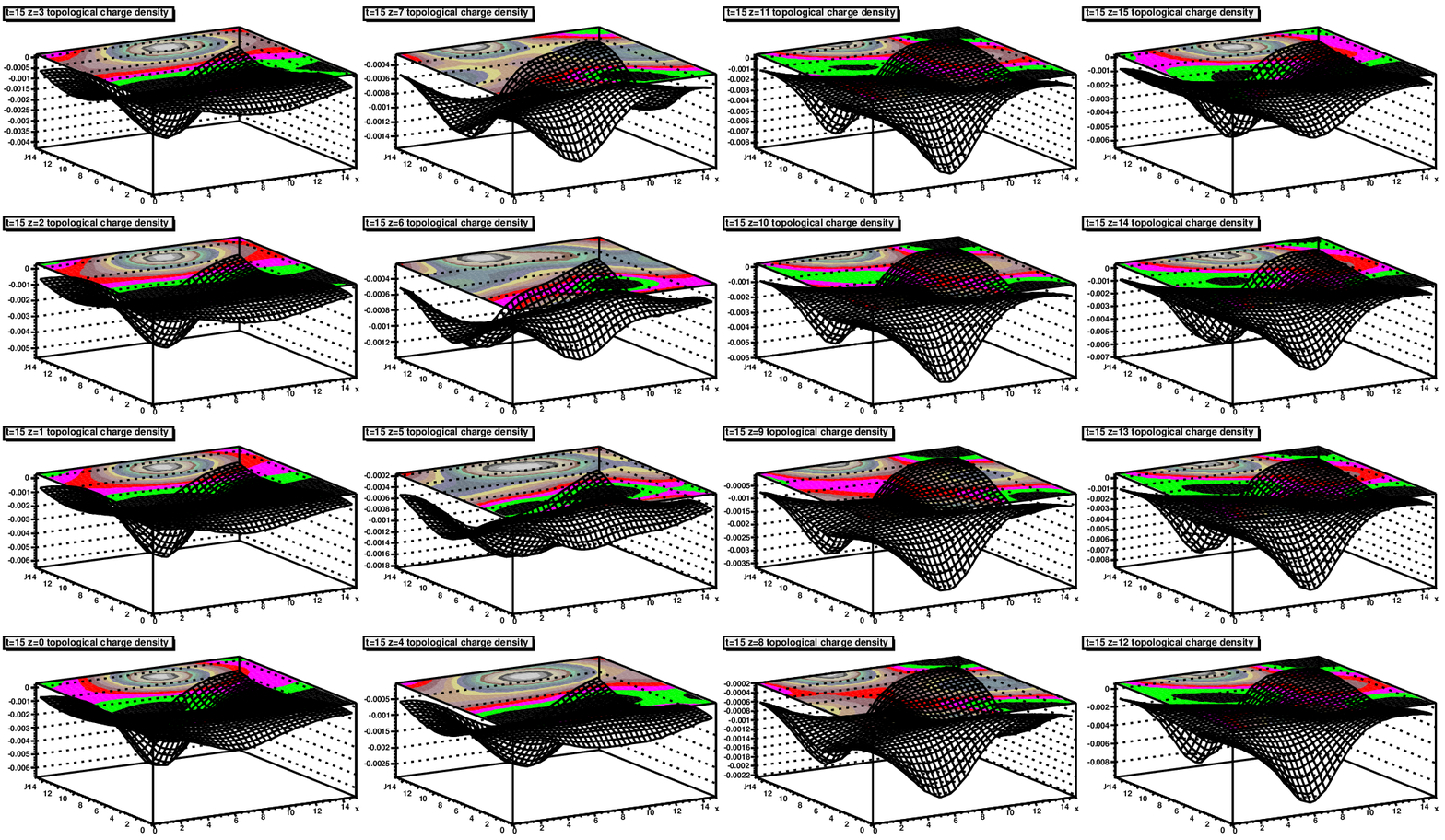}
 \caption{\label{lumps}Hyper surfaces with action density (top four rows) and topological 
charge density (bottom four rows) of a configuration obtained after 1500 cooling sweeps
from a $16^4$ lattice Monte Carlo configuration generated at $\beta=9.515$.}
\end{figure}

Indeed, the presence of such (localized) lumps can be identified directly in the 
configurations. Furthermore, the presence of topological lumps reflects itself also 
in the action density. This is depicted for a hypersurface in a typical example 
configuration in figure \ref{lumps}. The one-to-one correspondence between the lumps 
of action and topological charge is very well visible, which supports the interpretation 
of topologically stable structures.

To understand the structure of the lumps better, we have analyzed them for some 
configurations by a cluster finding algorithm (described and used, e. g., 
in \cite{Ilgenfritz:2007xu}). The number of clusters found depends strongly on the 
lower bound of the local topological charge density applied in order to distinguish between 
the (outside) vacuum and a cluster. Nonetheless, with the running lower bound adjusted 
suitably, all the clusters found by this procedure contained almost all of both the 
topological charge and the action. Surprisingly, in most cases the number of clusters 
was around half of the topological charge of the configuration. Therefore, multiple-charged 
clusters dominate the structure of configurations with $|Q|>1$.  Furthermore, the sizes of 
the clusters within one configuration varied by typically one order of magnitude.
These findings indicate a very interesting substructure of the topological excitations. 
It may be worthwhile to pursue this investigation further in the future.

\begin{figure}
 \includegraphics[width=\linewidth]{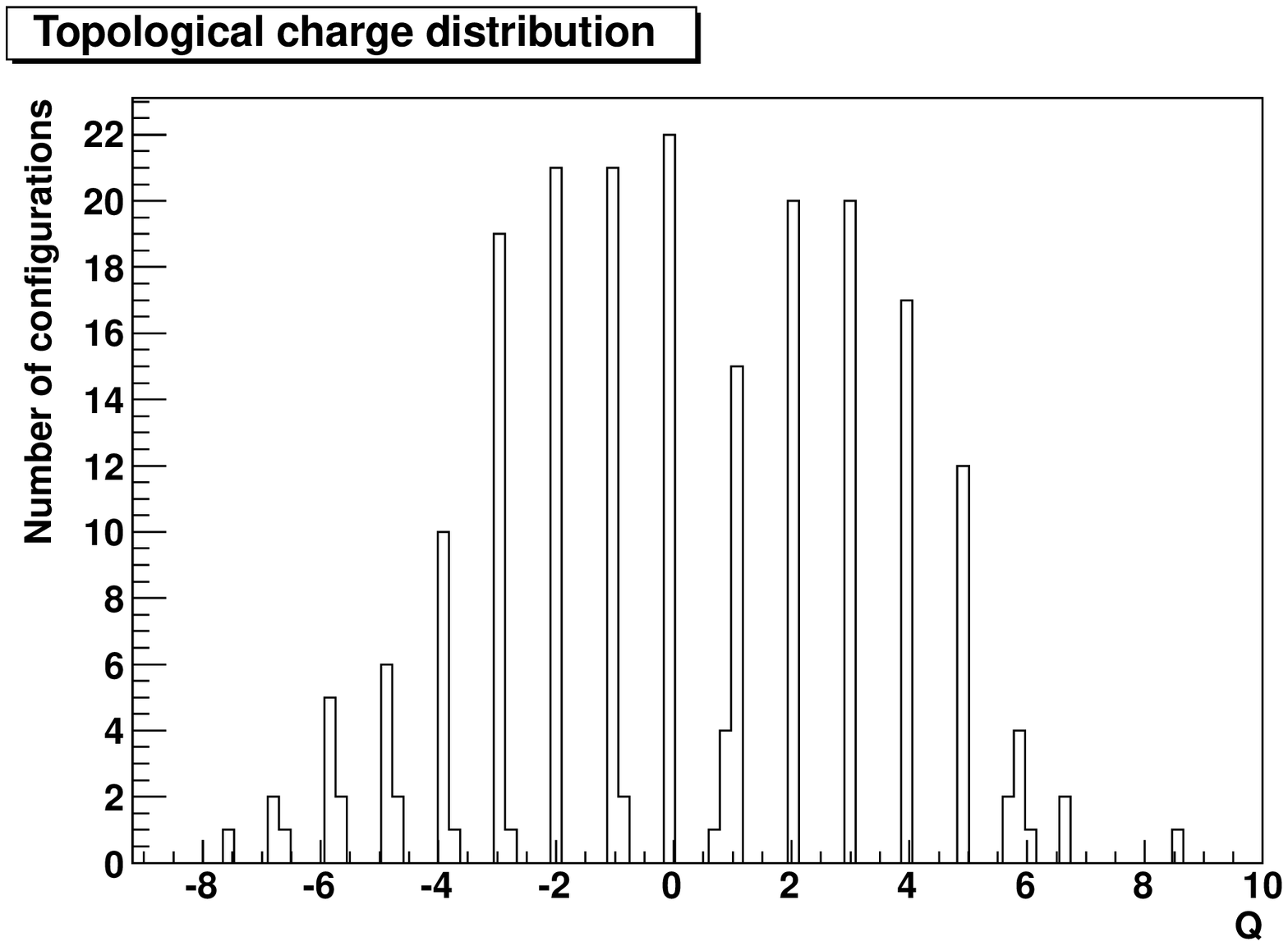}
 \caption{\label{qdis}Histogram of the topological charges per configuration after 1500 
sweeps on a $12^4$ lattice at $\beta=9.515$.}
\end{figure}

A typical distribution of the topological charge per configuration after 1500 cooling 
sweeps is shown in figure \ref{qdis}. It is clearly visible that the topological charge 
values are concentrated around integer values. The few intermediate values are likely 
from configurations which are in the process of stepping down from one plateau to a lower
one. Whether the observed distribution is Gaussian or follows a different multiplicity
distribution 
cannot be reliably determined with the available amount of data.
Nonetheless, the presence of (anti)selfdual topological lumps with integer topological 
charge is therefore well-established, as well as their correlation with action lumps.

It is possible to define a residual configuration, in which each link $U_\mu^r$ is given by
\be
U_\mu^r=U_\mu^{\text{cooled}-1}U_\mu\nn,
\ee
\no where $U_\mu$ is the original link and $U_\mu^{\text{cooled}}$ the cooled link. 
These ``residual configurations'' have an action which is almost independent of the 
cooling, and possess no discernible topological structures. They thus appear to be 
dominated by the ultraviolet fluctuations.

\section{Topological charge and susceptibility at zero and finite temperature}
\label{sfinitet}

After establishing the properties of the individual lumps, the next step is to determine 
their statistical properties forming full lattice configurations.
Their properties will be discussed first at zero temperature, 
to identify any kind of volume and discretization artifacts, and to give an estimate of 
the topological susceptibility in the continuum and infinite volume limit.

\begin{figure}
 \includegraphics[width=\linewidth]{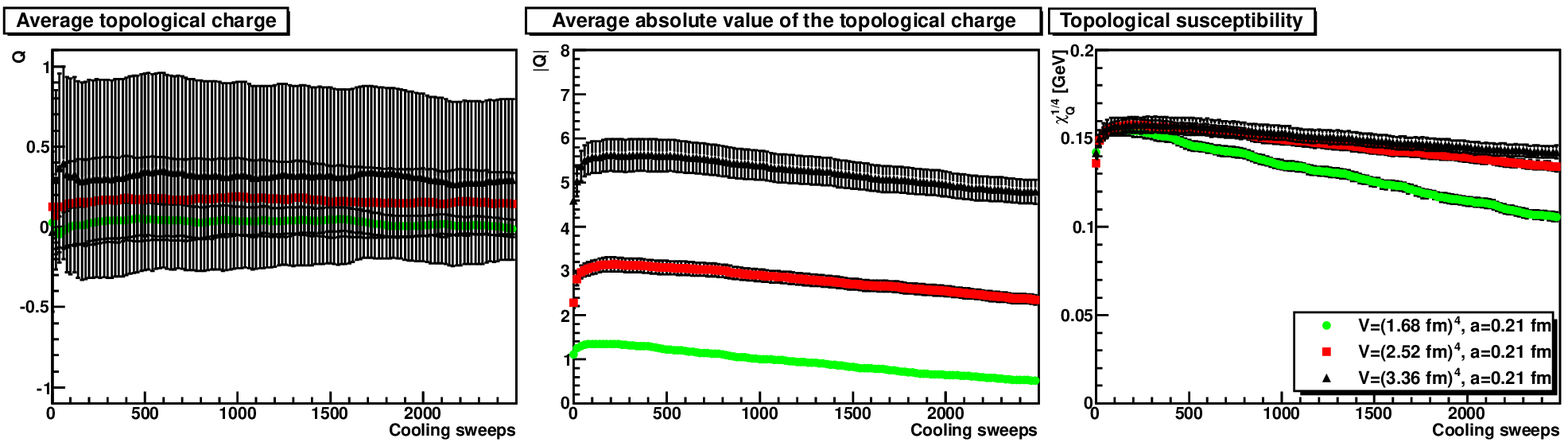}\\
 \includegraphics[width=\linewidth]{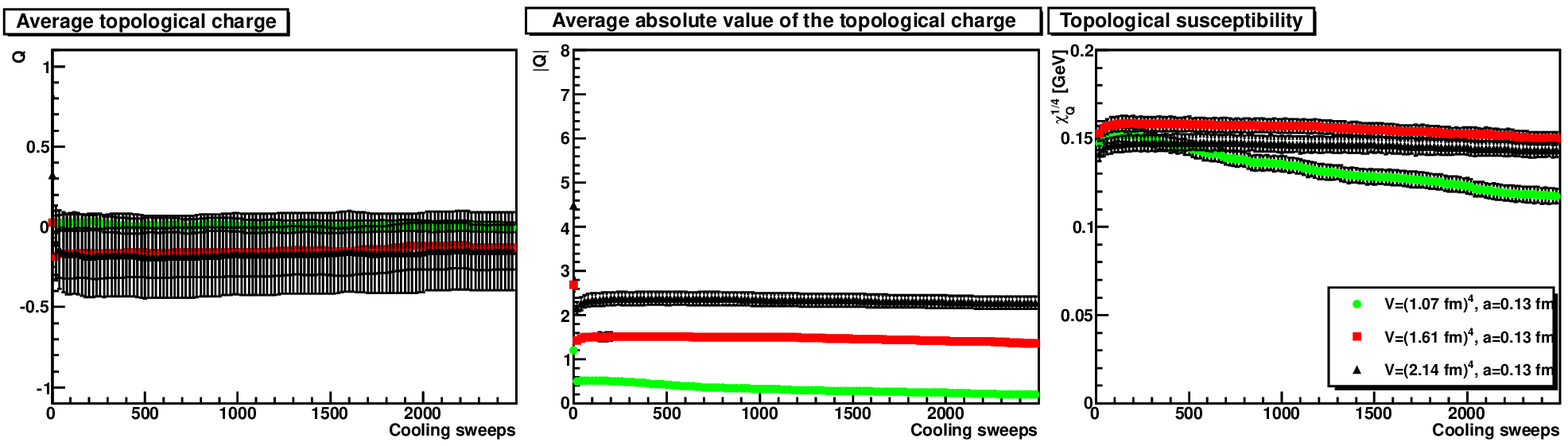}\\
 \includegraphics[width=\linewidth]{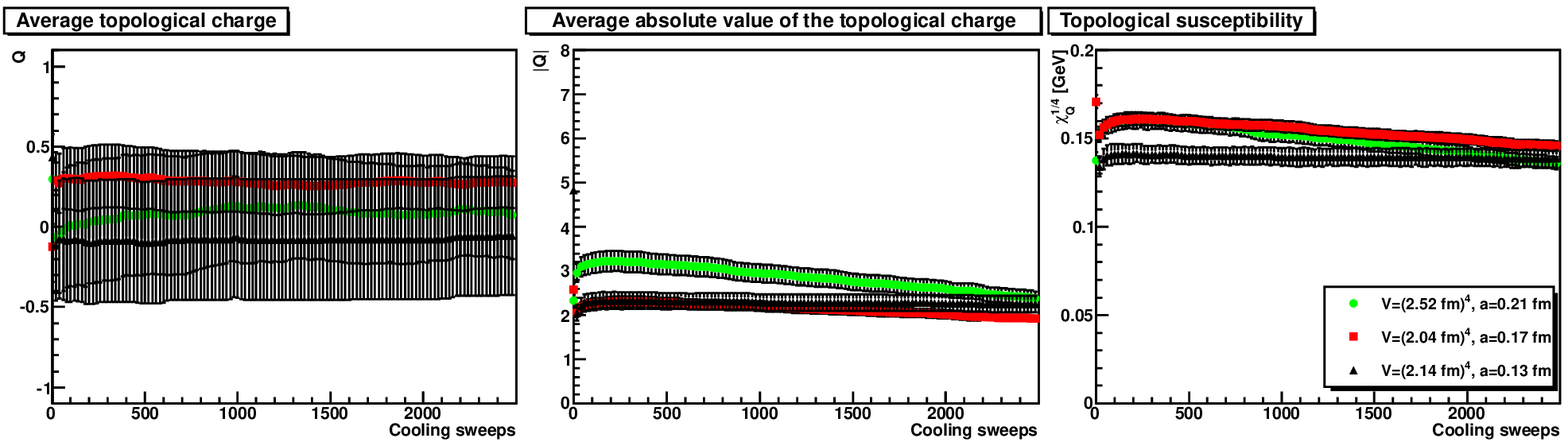}
 \caption{\label{x-x}The average of the topological charge, of the absolute value of the 
topological charge, and topological susceptibility as a function of the cooling sweeps 
for various volumes and discretizations. Only the measurements performed for every 20-th 
cooling sweep are shown.}
\end{figure}

In figure \ref{x-x}, the average and absolute values of the topological charge and the 
topological susceptibility are shown. As could already be inferred from figure \ref{qdis}, 
the average value of the topological charge is zero, though with rather large errors. 
The average absolute value of the topological charge increases quickly with volume. 
This indicates that with larger and larger volume more and more topological lumps fit into 
the given lattice volume. At the same time this number is rather insensitive to the lattice 
spacing. Thus, even with a rather coarse lattice the topological structure of the 
cooled vacuum 
seems to be well resolvable. Finally, the topological susceptibility turns out to be 
neither very sensitive to volume nor to the discretization. It also changes only weakly 
as a function of the number of cooling sweeps, and therefore appears to be a good 
observable. Its fourth root has a value of about 150 MeV, with a one sigma error band 
of the order of 25 MeV for all investigated cases. It is thus about six sigma away from 
zero, giving a rather good evidence for a non-zero topological susceptibility, 
apart from possible systematic errors. Thus, G$_2$ is in this respect rather similar, 
both qualitatively and quantitatively, to SU($N$) Yang-Mills theory.

Another feature of SU($N$) Yang-Mills theories is that the topological properties change 
at the phase transition. This is often invoked to explain both the restoration of chiral 
symmetry as well as deconfinement \cite{Schafer:1996wv,Greensite:2003bk}. Although G$_2$ 
Yang-Mills theory has no deconfinement comparable to QCD, it shows a sudden rise of the
Polyakov loop at some temperature, and chiral symmetry is
restored at the same transition temperature in a similar way as in usual 
QCD \cite{Danzer:2008bk}.
If the restoration of chiral symmetry is indeed related to the topological degrees of 
freedom, the phase transition should therefore reflect itself in the topological properties.

\begin{figure}
 \includegraphics[width=\linewidth]{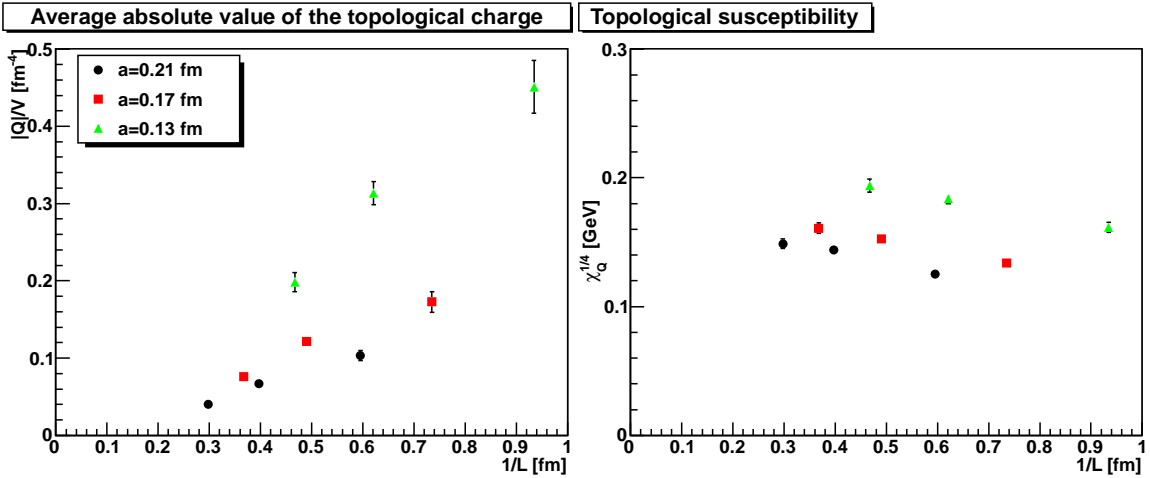}
 \caption{\label{vdep-x}Volume and discretization dependence of the average absolute 
value of the topological charge density $|Q|/V$ 
and of the fourth root of the topological susceptibility after 1500 cooling sweeps.}
\end{figure}

The investigation of this interconnection is somewhat complicated by the presence of a 
bulk transition on coarse lattices, requiring to use rather fine lattices with at least 
$N_t=6$ \cite{Cossu:2007dk}. This implies significant changes in the physical volumes, 
as going to larger volumes incurs too large computational costs. 

This has to be taken into account. To address it in an at least heuristic way, the 
volume-dependence of the average value of the absolute value of the topological charge 
and of the topological susceptibility after a fixed number of cooling sweeps
is shown in figure \ref{vdep-x}.
The obtained qualitative results do not depend on the number of cooling sweeps: 
The topological charge density $|Q|/V$ depends strongly on both volume and discretization. 
It appears that the larger the volume the less the absolute topological charge. At the same 
time, the better the discretization, the higher the absolute topological charge. Given that with larger lattice volumes more and more topological lumps of both signs should fit into the system, it appears likely that the total charge diminishes quickly with volume. Discretization effects seem to offset this to some extent. However, for a full understanding the detailed cluster structure must be understood more systematically, which requires significantly more resources. Notice in this context that cooling has a tendency to lower the action by eliminating different-sign topological lumps, which may affect this outcome. This should not as strongly affect the topological fluctuations, in agreement with figure \ref{vdep-x}.

This makes it rather complicated to disentangle the temperature and volume effects for 
the topological charge density, even if the topological charge turns out to be a decreasing 
function of temperature.

\begin{figure}
 \includegraphics[width=\linewidth]{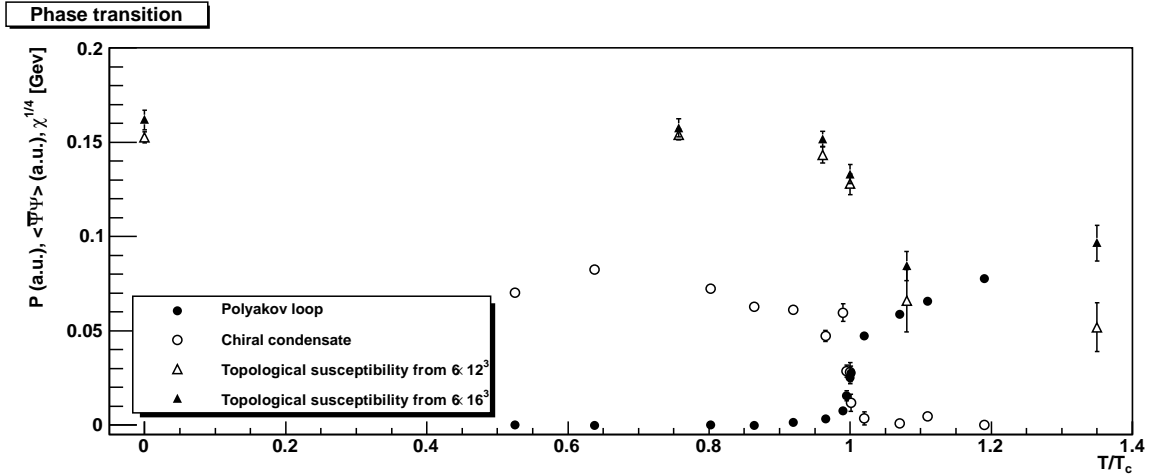}
 \caption{\label{pt}The topological susceptibility as a function of temperature on 
both the $6\times 12^3$ and the $6\times 16^3$ lattices. For comparison the results 
for the Polyakov loop and the chiral condensate are also shown, both 
from \cite{Danzer:2008bk}}
\end{figure}

From these observation it appears reasonable to use only the topological susceptibility 
to study the change of topological properties with temperature. It is shown in 
figure \ref{pt}, compared to the Polyakov loop and the chiral condensate. The topological 
susceptibility reacts to the phase transition by starting to drop from its zero-temperature 
value to a finite, high-temperature value. However, the drop is not as sharp as for the 
Polyakov loop and the chiral condensate, and the high-temperature value is 
reached somewhat above the critical temperature, at $T/T_c \approx 1.1$. 

Thus, the phase transition leaves an imprint on the topological properties of the theory. 
Interestingly, topological degrees of freedom are still present in the high-temperature 
phase, but a manual survey showed that only very few topological lumps remain. 
That the topological susceptibility is not reacting to temperature below the phase 
transition and does not vanish in the high-temperature phase 
close to the transition temperature, is similar to the case of SU($N$) gauge theory 
for not too large $N$ \cite{Gattringer:2002mr,DelDebbio:2004rw,Alles:1996nm}. 
In contrast to pure gluodynamics, in QCD the residual topological susceptibility in 
the high-temperature phase could be suppressed with an increasing number of quark 
flavors \cite{Alles:2000cg}. However, more recent investigations find a sharper drop for the Yang-Mills case, while the QCD transition is smoother \cite{Bornyakov:2012aa}. If this would be confirmed then G$_2$ Yang-Mills theory would behave more similar to the QCD case.

However, one should be wary that systematic effects, especially from both the definition 
of the topological charge operator and the cooling procedure, can substantially alter 
the result. 
E.\ g., for the $6\times 16^3$ lattice at the highest temperature the fourth root of the
topological susceptibility, $\chi^\frac{1}{4}$, changes from 0.099(6) over 0.093(7) to 
0.085(6) GeV when increasing the number of cooling sweeps from 500 over 1500 to 2500.

\section{Summary}\label{ssum}

We have presented the first numerical lattice investigation of topological properties of G$_2$ Yang-Mills theory. We found that topological lumps exist, which can be identified individually by cooling, and which provide an integer topological charge for a given configuration. From this, we could determine the topological susceptibility and its dependence on temperature. This susceptibility changes at the phase transition from one finite to another finite value. In total, the results show a remarkable qualitative, and to some extent even quantitative, similarity to SU($N$) Yang-Mills theories at small $N$. This underlines once more that, in spite of the group-theoretical differences, especially the trivial center, G$_2$ Yang-Mills theory is quite similar to SU($N$) Yang-Mills theory. This emphasizes that the center structure is for many quantities of little concern.\\

\no{\bf Acknowledgments}\\

We are grateful to Christof Gattringer for helpful discussions. A.\ M.\ was supported by the FWF under grant number M1099-N16 and by the DFG under grant number MA 3935/5-1. Computing time was provided by the HPC cluster of the University of Graz. The ROOT framework \cite{Brun:1997pa} has been used in this project.

\bibliographystyle{bibstyle}
\bibliography{bib}


\end{document}